\documentclass[twocolumn,showpacs,aps,prl,superscriptaddress]{revtex4}

\usepackage{graphicx}
\usepackage{dcolumn}
\usepackage{epsfig}
\usepackage{amsmath}
\usepackage{units}
\RequirePackage{xspace}
\usepackage{relsize}
\input{pubboard/babarsym.tex} 

\newcommand{\BABARPubYear}    {08}
\newcommand{\BABARPubNumber}  {053}

\newcommand{\SLACPubNumber} {13452}

\def\rhoz     {\ensuremath{\rho^0}\xspace}
\def\fz     {\ensuremath{f_0}\xspace}
\def\rhop   {\ensuremath{\rho^+}\xspace}
\def\Acp    {\ensuremath{{\cal A}_{\CP}}\xspace}
\def\rhom   {\ensuremath{\rho^-}\xspace}
\def\rfz    {\ensuremath{\rhop\fz}\xspace}
\def\btorr  {\ensuremath{\Bp\to\rhop\rhoz}\xspace}

\def\mrz    {\ensuremath{m_{\pip\pim}}\xspace}

\def\btorfz {\ensuremath{\Bp\to\rhop\fz}\xspace}
\def\rprz   {\ensuremath{\rhop\rhoz}\xspace}

\newcommand {\aeff} {\ensuremath{\alpha_{\rm eff}}\xspace}

\newcommand{\vub}{V_{ub}}

\newcommand{\vud}{V_{ud}}
\newcommand{\vtb}{V_{tb}}
\newcommand{\vtd}{V_{td}}

\newcommand{\etal}{{\it et. al.}}
\begin{document}

\begin{flushleft}
SLAC-PUB-\SLACPubNumber \\
\babar-PUB-\BABARPubYear/{\BABARPubNumber}\\
arXiv: 0901.3522 [hep-ex]
\end{flushleft}

\title{ \large \bf\boldmath Improved 
 Measurement of $B^+\to\rho^+\rho^0$ and Determination of the 
 Quark-Mixing Phase Angle~$\alpha$}

%
\author{B.~Aubert}
\author{Y.~Karyotakis}
\author{J.~P.~Lees}
\author{V.~Poireau}
\author{E.~Prencipe}
\author{X.~Prudent}
\author{V.~Tisserand}
\affiliation{Laboratoire d'Annecy-le-Vieux de Physique des Particules (LAPP), Université de Savoie, CNRS/IN2P3, F-74941 Annecy-Le-Vieux, France }
\author{J.~Garra~Tico}
\author{E.~Grauges}
\affiliation{Universitat de Barcelona, Facultat de Fisica, Departament ECM, E-08028 Barcelona, Spain }
\author{L.~Lopez$^{ab}$ }
\author{A.~Palano$^{ab}$ }
\author{M.~Pappagallo$^{ab}$ }
\affiliation{INFN Sezione di Bari$^{a}$; Dipartmento di Fisica, Universit\`a di Bari$^{b}$, I-70126 Bari, Italy }
% ...
\collaboration{The \babar\ Collaboration}
\noaffiliation

\begin{abstract}
  We present improved measurements of the 
  branching fraction ${\cal B}$, the longitudinal polarization fraction $f_L$, 
  and the direct \CP asymmetry \Acp in the \B meson decay channel
  $B^+\to\rho^+\rho^0$.
  The data sample was collected with the {\babar} detector at SLAC. 
  The results are ${\cal B}\,(\Bp\ra\rprz)=(23.7\pm1.4\pm1.4)\times10^{-6}$, 
  $f_L=0.950\pm0.015\pm0.006$, and $\Acp=-0.054\pm0.055\pm0.010$,
  where the uncertainties are statistical and systematic, respectively. 
  Based on these results, we perform an isospin 
  analysis and determine the CKM phase angle
  $\alpha=\mathrm{arg}\left(-{\vtd\vtb^*/\vud\vub^*}\right)$
  to be $(92.4^{+6.0}_{-6.5})^{\circ}$.
\end{abstract}

\pacs{13.25.Hw, 12.15.Hh, 11.30.Er}
\maketitle

In the Standard Model (SM),
the weak interaction couplings of quarks are described by elements 
$V_{ij}$ of the Cabibbo-Kobayashi-Maskawa (CKM) matrix~\cite{ckm},
where $i=u,c,t$ and $j=d,s,b$ are quark indices.
The CKM elements are complex,
introducing violation of charge-parity (\CP) symmetry.
Unitarity of the CKM matrix yields a relationship between
the $V_{ij}$ that can be represented as a triangle in the complex plane.
The SM mechanism for \CP violations can be tested
through measurement of the sides and angles of
this unitarity triangle (UT)~\cite{PDG}.
An approximate result \aeff for the UT angle
$\alpha=\mathrm{arg}\left(-{\vtd\vtb^*/\vud\vub^*}\right)$
can be obtained from \B meson decays to \CP eigenstates
dominated by tree-level $b\to u \bar ud$ amplitudes,
such as $\B\ra\rho\rho$ decays
(see, e.g., Refs.~\cite{PDG,ckmfitter}).
The correction $\Delta\alpha=\alpha-\aeff$,
which accounts for loop amplitudes,
can be extracted from an analysis
of the branching fractions and \CP asymmetries 
of the full set of isospin-related $b\to u \bar ud$ channels~\cite{alphaPRL}.
One of the most favorable methods to determine $\alpha$ is through an
isospin analysis of the $\B\ra\rho\rho$ system~\cite{PDG,ckmfitter}.

Here,
we present updated results for the $\Bp\ra\rhop\rhoz$ channel,
with $\rho^+\to\pip\piz$ and $\rho^0\to\pip\pim$,
leading to an improved determination of~$\alpha$.
Previous studies are presented
in Refs.~\cite{rho0rhop,bib-previous-rprz}.
We measure the branching fraction ${\cal B}$,
the longitudinal polarization fraction $f_L$,
and the direct \CP asymmetry
$\Acp\equiv(\Gamma_{B^-}-\Gamma_{B^+})/(\Gamma_{B^-}+\Gamma_{B^+})$,
with $\Gamma_{\Bpm}$ the \Bpm decay width.
Significant deviation of \Acp from the SM prediction of zero
could indicate new physics.
We also search for the as-yet-unobserved decay $\btorfz(980)$,
with $f_0\to\pip\pim$.
The use of charge conjugate reactions is implied throughout.

The analysis is based on $(465\pm5)\times 10^6$ \BB events 
(424~\invfb) collected on the \FourS resonance
[center-of-mass (CM) energy $\sqrt{s}=10.58~\gev$]
with the {\babar} detector~\cite{BABARNIM} at the \pep2
asymmetric energy \epem collider at SLAC.
Compared to our previous study~\cite{rho0rhop},
the analysis incorporates higher signal efficiency
and background rejection, twice as much data,
and improved procedures to reconstruct charged particles
and to account for correlations in the backgrounds.
Simulated event samples based on Monte Carlo (MC) event generation
are used to determine signal and background characteristics,
optimize selection criteria, and evaluate efficiencies.

$\Bp\ra\rho^+\rho^0$ decays are described
by a superposition of two transversely
(helicity $\pm 1$) and one longitudinally (helicity 0)
polarized amplitudes.
Our acceptance is independent of the angle between the 
two $\rho$ decay planes in the \B rest frame.
We integrate over this angle to obtain an expression for
$(1/\Gamma)\,d^2\Gamma/\left(d\cos\theta_{\rho^0} d\cos\theta_{\rho^+}\right)$:
\begin{eqnarray}
{9\over16}\left[4f_L\cos^2\theta_{\rho^0}\cos^2\theta_{\rho^+}
   + (1-f_L)\sin^2\theta_{\rho^0}\sin^2\theta_{\rho^+}\right],
\label{eq:2D}
\end{eqnarray}
with $f_L\equiv\Gamma_L/\Gamma$,
where $\Gamma$ is the total decay width,
$\Gamma_L$ is the partial width to the longitudinally-polarized mode,
and the $\rho^0$ ($\rho^+$) helicity angle 
$\theta_{\rho^0}$ ($\theta_{\rho^+}$)
is the angle between the daughter $\pi^+$
in the $\rho^0$ ($\rho^+$) rest frame and
the direction of the boost from the \Bp rest frame.

A \B meson candidate is kinematically characterized
by the beam-energy-substituted mass
$\mes\equiv\sqrt{s/4-(p_B^*c)^2}/c^2$
and energy difference
$\DeltaE\equiv E_B^*-\sqrt{s}/2$,
where $E_B^*$ and $p_B^*$ are the CM
energy and momentum of the $B$ candidate, respectively.
Signal events peak at the nominal \B mass for \mes
and at zero for \DeltaE,
with resolutions of 3~\mevcc and 30~MeV, respectively.

The \piz mesons are reconstructed through $\piz\ra\g\g$.
The \g is required to be consistent with a
single electromagnetic shower.
The \g and \piz laboratory energies must be larger 
than 30~MeV and 0.2~GeV, respectively.  
The mass of a \piz candidate (resolution 6~\mevcc)
is required to lie within [0.115, 0.150]~\gevcc and
is subsequently constrained
to its nominal value~\cite{PDG}.

The \piz (\pim) candidate is combined with a
\pip to form a $\rho^+$ $(\rho^0)$.
The \pipm are identified
with measurements of specific energy loss in the tracking chambers, 
and radiation angles and photon multiplicity
in a ring-imaging Cherenkov detector~\cite{BABARNIM}.
The $\rho^+(\rho^0)$ candidate mass 
$m_{\pip\piz}$ ($m_{\pip\pim}$) must lie
within [0.52, 1.06]~\gevcc.  
$\rho^+$ candidates with mis-reconstructed \piz mesons tend to 
cluster near \mbox{$\cos\theta_{\rhop}\approx 1$},
so we require \mbox{$\cos\theta_{\rhop}\le0.8$.}
The \Bp candidates must satisfy
$5.26<\mes<5.29$~\gevcc and $|\DeltaE|<0.15$~GeV.
In cases of multiple \Bp candidates (about 10\% of events),
the candidate with the largest \Bp vertex~\cite{BTagger}
fit probability is retained.

Background from $\Bb$$\,\ra\,$$\Db^{(*)}X$ decays,
due to $\Dzb$$\,\ra\,$$\Kp\pim(\pi^0)$ with kaon misidentification
and $\Dzb\to\pip\pim\pi^0$, 
is suppressed by requiring
the $K^+\pi^-(\pi^0$) or $\pi^+\pi^-\pi^0$ invariant mass
to lie outside $\pm4\,\sigma$ of the nominal $\Dz$ mass~\cite{PDG},
with $\sigma\approx 9$~\mevcc
the \Dz mass resolution.

The dominant background,
from random combinations of particles in continuum events
($\epem\to\qqbar$, with $q=u,d,s,c$),
is suppressed by requiring $|\cos\theta_T|<0.8$~\cite{bib-babar-2004a}, 
with $\theta_T$ the angle between the thrust axis of the \B candidate's
decay products and the thrust axis of the remaining 
particles in the event (ROE), 
evaluated in the CM frame,
and by employing a neural network algorithm
based on 11 variables calculated in the CM:
$|\cos\theta_T|$;
the cosines of the angles with respect to the beam axis of
the \B momentum and \B thrust axis
(we use the absolute value for the latter variable);
the momentum-weighted sums $L_0$ and $L_2$~\cite{bib-babar-2004a},
determined with charged and neutral particles separately;
the sum of transverse momenta of the ROE particles
with respect to the beam axis;
the ratio of the second to zeroth Fox-Wolfram moments~\cite{R2ref};
the proper time difference between the \B 
and \Bb candidates divided by its uncertainty; 
and \B-tagging information from ROE particles~\cite{BTagger}. 
The neural network output $NN$ peaks near 0 and 1 for continuum
and signal events, respectively.
We require $NN>0.2$,
which rejects about 5\% of the signal and 60\% of the continuum events.

We examine the remaining \B backgrounds and
identify nine channels with peaking structures in \mes or \DeltaE
that can potentially mimic signal events:
$\Bp\to\pi^0  a_1^+(1260)$,
$\pi^+  a_1^0$, $\rho^0\pi^+\pi^0$, $\rho^+\pi^+\pi^-$,
$\rho^-\pi^+\pi^+$, $\pi^0\pi^-\pi^+\pi^+$, $\omega\rho^+$,
$f_0\pi^0\pi^+$, and $\eta'\rho^+$, 
with $a_1\to\rho\pi$, $\omega\to\pip\pim$, $f_0\to\pip\pim$, 
and $\eta'\to\gamma\rhoz$.
All other \B backgrounds
are combined into a ``non-peaking'' \BB\ background component.

An extended unbinned maximum likelihood (ML) fit
is applied to the selected events. 
The fit has 14 components: 
signal $\rho^+\rho^0$ events, 
taken to be $\Bp\ra\rho^+\rho^0$ events that are correctly reconstructed;
self-cross-feed (SxF) events,
defined as mis-reconstructed $\Bp\ra\rho^+\rho^0$ events
(29\% of the $\Bp\ra\rho^+\rho^0$ sample);
signal $\Bp\ra\rho^+f_0$ events,
including both correctly and incorrectly reconstructed 
events to increase efficiency;
non-peaking \BB background;
continuum background;
and the nine peaking \BB background channels listed above. 
The $\rho^+\rho^0$ signal and SxF components are further divided into
categories with either longitudinal or transverse \mbox{polarization.}

The likelihood function is
$  
{\cal L} =(1/ N!)\exp{\left(-\sum_{j} n_j\right)} 
$
$
 \prod_{i=1}^N \left[\sum_{j} n_j  {\cal P}_j \left({\bf x}_i\right)\right],
$
with $N$ the number of events,
$n_j$ the yield of component $j$, 
${\cal P}_j ({\bf x}_i)$ the probability density function (PDF) 
for event $i$ to be associated with component~$j$,
and ${\bf x}_i$
the seven experimental observables specified
in Eq.~(\ref{PDFs6}) below.
The signal $\rho^+\rho^0$, $\rho^+f_0$,
continuum and non-peaking \BB background yields are allowed to vary in the fit.
The $\rhop\rhoz$ SxF yield is fixed to its expected value
based on the MC prediction for the SxF rate and the $\Bp\ra\rho^+\rho^0$
branching fraction determined here
(we iterate the fit to find this result).
The relative contributions of the $\rho^+\rho^0$ longitudinal 
and transverse polarization components are determined 
by allowing $f_L$ to vary,
with $f_L$ common to the signal and SxF events.
The three $\rho\pi\pi$ yields 
are varied under the requirement that they 
have the same branching fraction.
The $\pi^0  a_1^+$, $\pi^+  a_1^0$, $\omega\rho^+$, 
and $\eta'\rho^+$ yields are fixed according to their 
known branching fractions~\cite{PDG}.
The $\pi^0\pi^-\pi^+\pi^+$ and $f_0\pi^0\pi^+$ yields
are fixed assuming their branching fractions to be $10^{-5}$,
consistent with or larger than the
limits~\cite{rho0rho0,bib-belle-pipipipi}
for $\Bz\ra\pi^+\pi^-\pi^+\pi^+$ and~$f_0\,\pi^+\pi^-$ decays.

About 85\% of continuum events,
and 90\% of non-peaking \BB background events,
contain at least one mis-reconstructed $\rho$.
For these events,
we  find correlations of order 10\% between 
the $NN$, $m_{\pi\pi}$, and $\cos\theta_{\rho}$ variables,
and 
--~to account for these correlations~--
construct three-dimensional (3D) PDF's of the five variables
based on conditional PDF's ${\cal P}(x|y)$ of variable $x$ given the 
value of variable $y$:
${\cal P}_{3D} =  [{\cal P}(m_{\pip\pim}|\cos\theta_{\rho^0})
  \times{\cal P}(\cos\theta_{\rho^0}|NN)] 
  \times  [{\cal P}(m_{\pip\piz}|\cos\theta_{\rho^+})
  \times{\cal P}(\cos\theta_{\rho^+}|NN)]
  \times  {\cal P}(NN)$.
For example,
${\cal P}(m_{\pip\piz}|\cos\theta_{\rho^+})$ is constructed
by examining the $m_{\pip\piz}$ distribution in 
nine bins of $\cos\theta_{\rho^+}$,
fitting a second order polynomial to each bin,
and parameterizing how the coefficients of the polynomial 
vary between bins.
The fraction of events with a correctly reconstructed 
$\rho^+$ and $\rho^0$ is fixed to 
the MC prediction for the non-peaking \BB background and
allowed to vary for the continuum background.
For all other components,
the overall PDF's are defined as the product of seven 1D PDF's,
one for each observable.
The PDF's of the $\rho^+\rho^0$ signal and SxF helicity angles
take the form of Eq.~(\ref{eq:2D}),
with detector resolution and acceptance incorporated,
by summing the longitudinal ($L$) and transverse ($T$)
components with a relative fraction 
${f_L\epsilon_L/[f_L\epsilon_L+(1-f_L)\epsilon_T}]$,
with $\epsilon_L$ and $\epsilon_T$ the respective
reconstruction efficiencies,
leading to an effective 2D PDF in
$\cos\theta_{\rho^+}$ and $|\cos\theta_{\rho^0}|$:
\begin{eqnarray} 
  {\cal P}_j \left({\bf x}_i\right)
   &=&{\cal P}_j(\mes^i)~{\cal P}_j(\Delta E^i)
  {\cal P}_j({NN}^i)~{\cal P}_j(m^i_{\pi^+\pi^0})\nonumber\\ 
  &&\times{\cal P}_j(m^i_{\pi^+\pi^-})
  ~{\cal P}_j(\cos\theta^i_{\rho^+},~|\cos\theta^i_{\rho^0}|) .
\label{PDFs6}
\end{eqnarray}

The continuum background \mes and \DeltaE PDF's 
are derived from a 44~fb$^{-1}$ data sample collected 
40~MeV below the $\Upsilon(4S)$ mass. 
All other PDF's are derived from simulation. 
For \mes, the PDF's of signal and continuum are
parameterized by a Crystal Ball~\cite{CBALL}
and an ARGUS function~\cite{ARGUS}, respectively.
A relativistic Breit-Wigner function with a 
$p$-wave Blatt-Weisskopf form factor is used for
the $m_{\pi\pi}$ distributions in $\rho^+\rho^0$ signal events.
For the background, 
$m_{\pi\pi}$ is modeled by a combination of
a polynomial and the signal function.  
Slowly varying distributions
($\DeltaE$ for non-peaking backgrounds, and $\cos\theta_\rho$) 
are modeled by polynomials.
High statistics histograms are used 
for the $NN$ distributions.
The remaining variables are parameterized with sums of Gaussians,
e.g.,
the $m_{\pi\pi}$ distribution in \fz decays is modeled with a sum
of three Gaussians.
A large data control sample of 
$\Bp\to \Dzb\pi^+$ ($\Dzb\to \KS\pi^0$, $\KS\ra\pip\pim$) 
events is used to verify that the resolution and peak position of
the signal $\mes$ and $\DeltaE$ PDF's are accurately simulated.

\begin{figure}[t]
\begin{tabular}{c}
\includegraphics[width=0.24\textwidth]{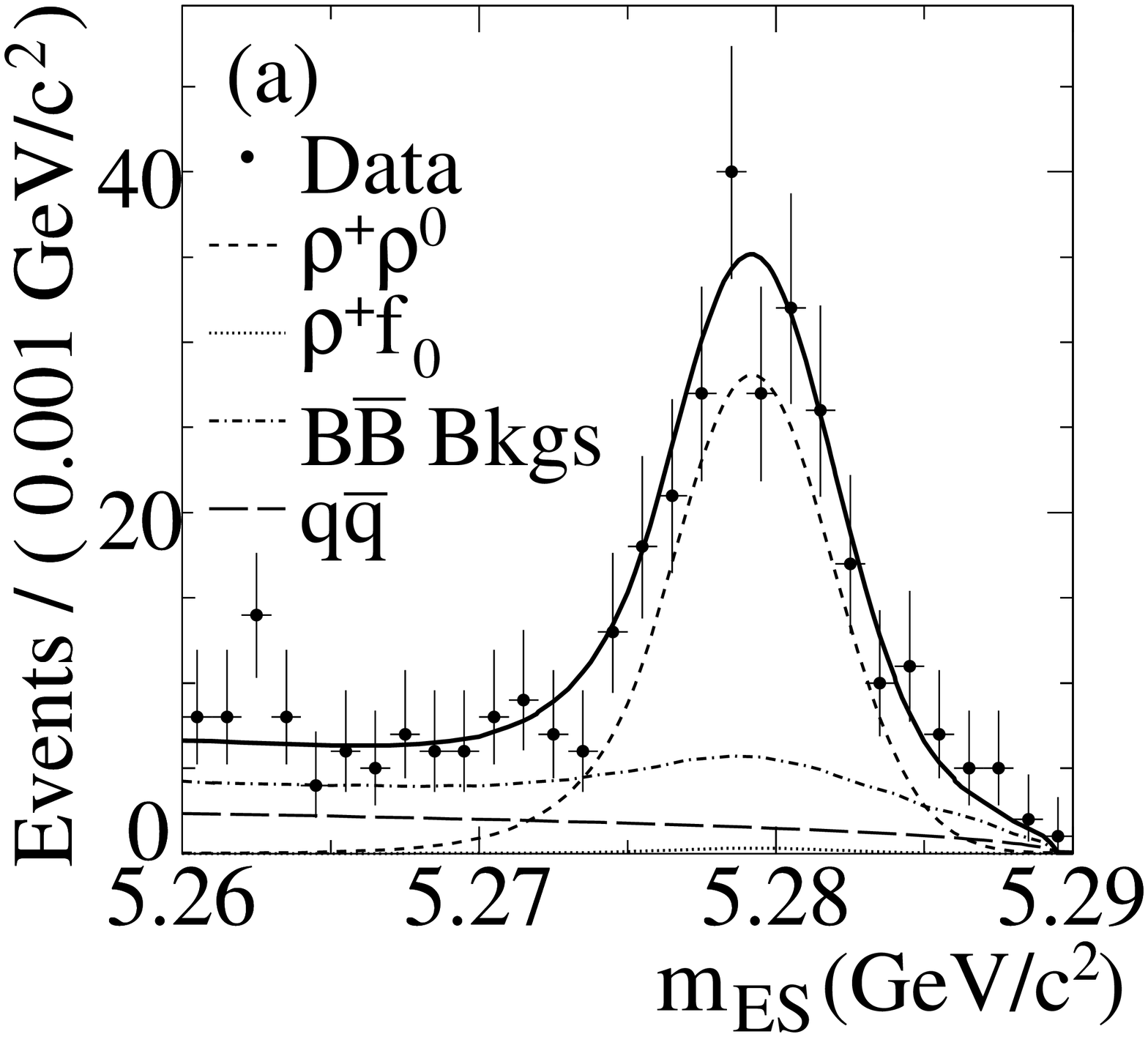}
 \includegraphics[width=0.24\textwidth]{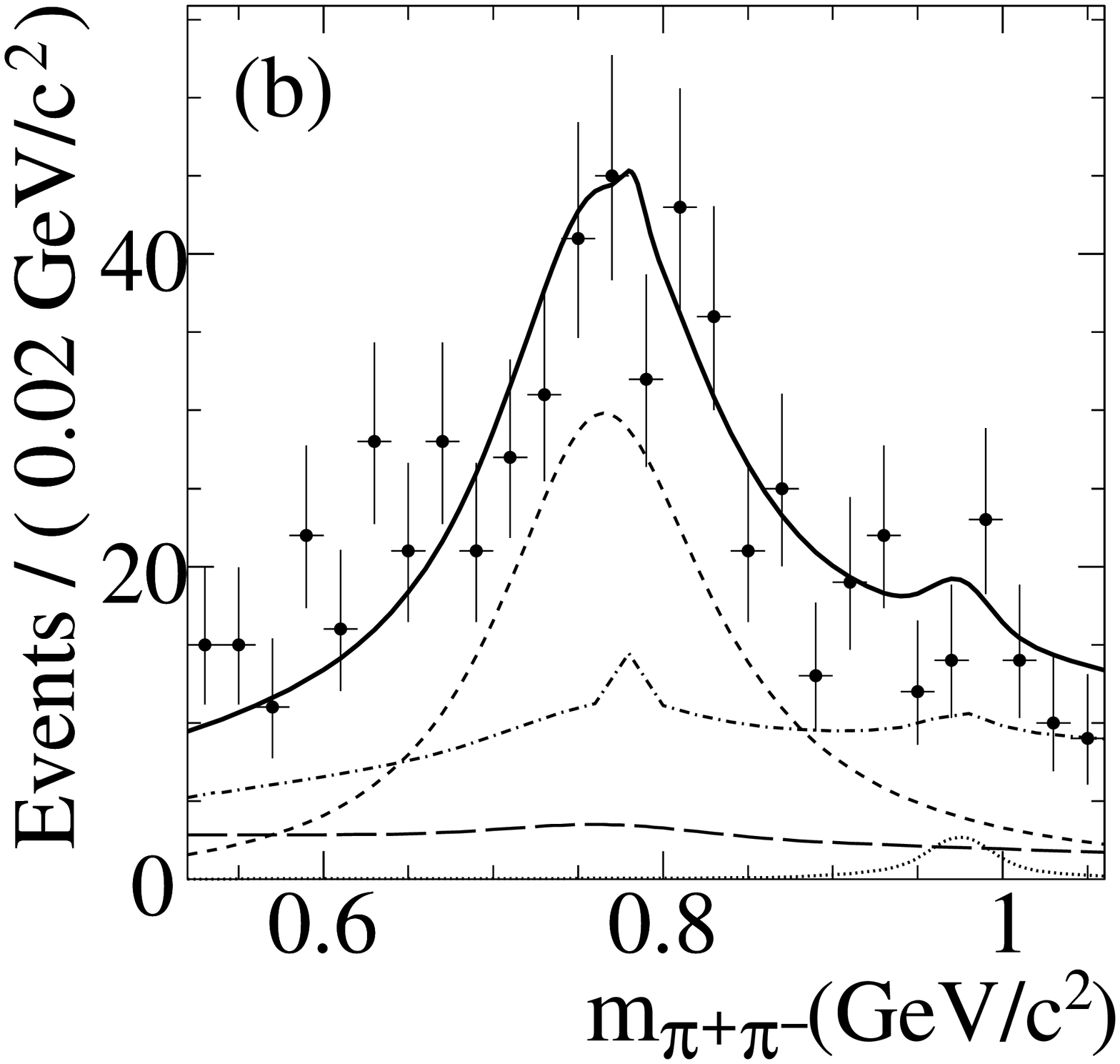}
\end{tabular}
\caption{Projections of the fit (solid curve) onto the (a) \mes 
and (b) $m_{\pi^+\pi^-}$ variables.
A requirement on the likelihood ratio that retains 38\% of the signal,
0.1\% of the continuum background,
and 1.3\% of the \BB background has been applied.
The peak in the \BB background at $m_{\pi^+\pi^-}\approx0.78$~GeV/$c^2$ 
is from  $\Bp\ra\rho^+\omega$ events with $\omega\ra\pip\pim$.
} 
\label{projrho0rho}
\end{figure}

The fit is applied to the sample of 82,224 selected events.
We allow 11 parameters to vary in the fit:
five parameters of continuum background PDF's, $f_L$,
and five yields as mentioned above.
We find $1122\pm 63$ (stat.)\!\! $\rhop\rhoz$ signal events,   
$50\pm 30$ (stat.)\! $\rhop\fz$ events, 
and $f_L=0.945\pm0.015$ (stat.).
The fit provides a simultaneous determination of the
number of $\Bp\ra\rhop\rhoz$ and  $\Bm\ra\rhom\rhoz$ signal events.
These fitted yields are used to determine  $\Acp=-0.054\pm0.055$~(stat.).
Fig.~\ref{projrho0rho} shows projections 
of the \mes and \mrz distributions.
To enhance the visibility of the signal, 
events are required to satisfy 
${\cal L}_i(S)/[{\cal L}_i(S)+{\cal L}_i(B)]>0.98$,  
where ${\cal L}_i(S)$ is the sum of the likelihood functions
for $\rho^+\rho^0$ and $\rhop\fz$ signal events
excluding the PDF of the plotted variable~$i$,
and ${\cal L}_i(B)$ is the corresponding sum of all other components. 

A possible bias, from unmodeled correlations,
is evaluated by applying the ML fit
to an ensemble of simulated experiments,
where the numbers of signal and background events in each
component correspond to those observed or fixed in the fit to data.
The continuum events are drawn from the PDF's,
while events for all other components
are drawn from MC samples.
The biases are determined to be
$71\pm3$ and $-31\pm1$ events for the 
signal $\rho^+\rho^0$ and $\rho^+f_0$ yields, 
and $-0.005\pm0.001$ for $f_L$,  
where the uncertainties are statistical.
The signal yields and $f_L$ are then corrected
by subtracting these biases.

The branching fractions are given by
the bias-corrected yields divided by the reconstruction efficiencies
and initial number of \BB pairs $N_{\BB}$.
From the simulations, 
the $\rho^+\rho^0$ signal efficiencies
including the \piz daughter branching fraction~\cite{PDG}
are $\epsilon_L$=[9.12$\pm$0.02~(stat.)]\% 
and $\epsilon_T$=[17.45$\pm$0.03~(stat.)]\%.
The corresponding result for $\rfz$ is [14.20$\pm$0.08~(stat.)]\%.
We assume that the \FourS decays to each of 
$\Bp\Bm$ and~$\Bz\Bzb$ 50\% of the~time.

The principal systematic uncertainties associated
with the ML fit are listed in Table~\ref{tab-additive}.
Uncertainties from the fit biases 
are defined by the quadratic sum of half the biases themselves
(for $f_L$, the full bias)
and the statistical uncertainties of the biases.
The uncertainties related to the signal and non-peaking \BB
background PDF's are assessed by varying the PDF parameters 
within their uncertainties.
For the signal,
the uncertainties of the PDF parameters are determined from
the $\Bp\to \Dzb\pip$ data control sample.
Variations of the 
$\pi^0  a_1^+$, $\pi^+  a_1^0$, $\omega\rho^+$, and $\eta'\rho^+$
branching fractions within their measured uncertainties,
and of the assumed $\pi^+\pi^-\pi^+\pi^0$ and $f_0\pi^+\piz$ 
branching fractions by $\pm$100\%,
define the systematic uncertainty associated
with the peaking \BB background.
The uncertainty associated with the SxF fraction is assessed
by varying the fixed SxF yield by $\pm10$\%. 
The other principal sources of systematic uncertainty are
the \piz reconstruction efficiency (3.0\%),
the track reconstruction efficiency (1.1\%), 
the \pipm identification efficiency (1.5\%), 
the uncertainty of $N_{\BB}$ (1.1\%),
and the selection requirements on $|\cos\theta_T|$ (1.0\%).
The individual terms are added in quadrature to define the 
total systematic uncertainties.

We find
${\cal B}(\btorr)=(23.7\pm1.4\pm1.4)\times10^{-6}$, 
$f_L=0.950\pm0.015\pm0.006$,
$\Acp=-0.054\pm0.055\pm0.010$,
and  ${\cal B}(B^+\to\rho^+f_0)\times{\cal B}(f_0\to\pip\pim)
=(1.21\pm0.44\pm0.40)\times10^{-6}$,
where the first (second) uncertainty is statistical (systematic).  
The ${\cal B}(\rhop\rhoz)$ result is larger than in Ref.~\cite{rho0rhop},
primarily because of the improved method used here
to account for correlations in the backgrounds.
The significance of the ${\cal B}(\rhop\fz)$ result
without (with) systematics
is 3.2 (2.2)~standard deviations.
We find $-0.15<\Acp<0.04$ and
${\cal B}(B^+\to\rho^+f_0)\times{\cal B}(f_0\to\pip\pim)<2.0\times10^{-6}$,
where these latter results correspond to 
the 90\% confidence level (CL) including systematics. 

We perform an isospin analysis of $B\to\rho\rho$ decays by
minimizing a $\chi^2$ that includes the measured quantities
expressed as the lengths of the sides of the \B and \Bb
isospin triangles~\cite{alphaPRL}.
We use the $\Bp\to\rho^+\rho^0$ branching fraction and $f_L$ results
presented here,
with the branching fractions, polarizations, and
\CP-violating parameters 
in $\Bz\to\rho^+\rho^-$~\cite{rhoprhom}
and $\Bz\to\rho^0\rho^0$~\cite{rho0rho0} decays.
We assume the uncertainties to be Gaussian-distributed and
neglect potential isospin $I=1$
and electroweak-loop amplitudes,
which are expected to be small~\cite{ckmfitter}.

\begin{table}[t]
\caption{Principal systematic uncertainties associated with the ML fit
(in events for the $\rho^+\rho^0$ and $\rhop\fz$ yields).}
\begin{tabular}{ccccc}
\hline
\hline
                         & $\rho^+\rho^0$ yield  & $\rhop\fz$ yield & $f_L$  & \Acp \\
\hline
Fit biases               &     35.5     &      15.3      & 0.005  & 0.001  \\
Signal PDF's             &     19.4     &      3.0       & 0.001  & 0.002 \\
Non-peaking \BB PDF's 
                         &     7.3      &      2.1       & 0.001  & 0.001 \\
Peaking \BB yields       &     16.3     &      21.1      & 0.003  & 0.001 \\ 
SxF fraction             &     7.9      &      0.1       & 0.001  & 0.001 \\
\hline
\hline
\end{tabular}
\label{tab-additive}
\end{table}

The CKM phase angle $\alpha$ and it correction 
$\Delta\alpha$ are found to be
$\alpha=(92.4^{+6.0}_{-6.5})^{\circ}$
and $-1.8^{\circ}<\Delta\alpha<6.7^{\circ}$ at 68\% CL,
significant improvements~\cite{bib-epaps}
compared to $\alpha=(82.6^{+32.6}_{-6.3})^\circ$ 
and $|\Delta\alpha|<15.7^\circ$~\cite{rho0rho0}
obtained with the same $\rho^+\rho^-$ and $\rho^0\rho^0$ measurements,
but the previous $\Bp\to\rho^+\rho^0$ results~\cite{rho0rhop},
or $\alpha=(91.7\pm14.9)^{\circ}$
from the Belle Collaboration~\cite{bib-belle-pipipipi}.
The improvement is primarily due to the increase
in ${\cal B}(\rhop\rhoz)$ compared to our previous result.
${\cal B}(\rhop\rhoz)$
determines the length of the common base of the 
isospin triangles for the \B and \Bb decays.
The increase in the base length flattens both triangles,
making the four possible solutions~\cite{alphaPRL} nearly degenerate.

In summary, 
we have improved the precision of the measurements of the
$\Bp\to\rho^+\rho^0$ decay branching
and longitudinal polarization fractions,
leading to a significant improvement in the determination of
the CKM phase angle~$\alpha$ 
based on the favored $\B\ra\rho\rho$ isospin method.
We set a 90\% CL upper limit of $2.0\times10^{-6}$ on the
branching fraction of $B^+\to\rho^+f_0(980)$ with $f_0\ra\pip\pim$.

We are grateful for the excellent luminosity and machine conditions
provided by our \pep2\ colleagues, 
and for the substantial dedicated effort from
the computing organizations that support \babar.
The collaborating institutions wish to thank 
SLAC for its support and kind hospitality. 
This work is supported by
DOE
and NSF (USA),
NSERC (Canada),
CEA and
CNRS-IN2P3
(France),
BMBF and DFG
(Germany),
INFN (Italy),
FOM (The Netherlands),
NFR (Norway),
MES (Russia),
MEC (Spain), and
STFC (United Kingdom). 
Individuals have received support from the
Marie Curie EIF (European Union) and
the A.~P.~Sloan Foundation.

\end{document}